\documentclass[12pt,a4paper]{article}
\usepackage{graphicx}
\usepackage{abstract}
\usepackage{amsmath}
\usepackage[left=1in, right=1in, top=1in, bottom=1in]{geometry}
\usepackage[dvipsnames]{xcolor} 
\usepackage{soul}
\usepackage{braket}
\usepackage{authblk}
\usepackage{placeins} 
\usepackage{caption}
\usepackage{soul}
\usepackage{caption}
\usepackage{subcaption}
\usepackage{float}
\usepackage{placeins}
\usepackage{geometry}
\usepackage{chngcntr}

\title{\textbf{Reducing the number of qubits in  quantum simulations of one dimensional many-body Hamiltonians}}
\author[1]{Somayeh Mehrabankar\thanks{Corresponding author: somayeh.mehrabankar@yahoo.com, mehso@uv.es}}
\author[2]{Miguel Ángel García-March}
\author[3]{Carmen G. Almudéver}
\author[1]{Armando Pérez}
\affil[1]{Departamento de Física Teórica and IFIC, Universidad de Valencia-CSIC, 
46100 Burjassot (Valencia), Spain}
\affil[2]{Instituto Universitario de Matemática Pura y Aplicada, Universitat Politècnica de València, 46022 València, Spain}
\affil[3]{Departamento de Informática de Sistemas y Computadores, Universitat Politècnica de València, 46022, València,Spain}

\date{}  
\begin{document}
\maketitle

\begin{abstract}    
We investigate the Ising and Heisenberg models using the Block Renormalization Group Method (BRGM), focusing on its behavior across different system sizes. The BRGM reduces the number of spins by a factor of 1/2 (1/3) for the Ising (Heisenberg) model, effectively preserving essential physical features of the model while using only a fraction of the spins. Through a comparative analysis, we demonstrate that as the system size increases, there is an exponential convergence between results obtained from the original and renormalized Ising Hamiltonians, provided the coupling constants are redefined accordingly. Remarkably, for a spin chain with 24 spins, all physical features, including magnetization, correlation function, and entanglement entropy, exhibit an exact correspondence with the results from the original Hamiltonian.  The study of the Heisenberg model also shows this tendency, although complete convergence may appear for a size much larger than 24 spins, and is therefore beyond our computational capabilities. The success of BRGM in accurately characterizing the Ising model, even with a relatively small number of spins, underscores its robustness and utility in studying complex physical systems, and facilitates its simulation on current NISQ computers, where the available number of qubits is largely constrained.
\end{abstract}
\newpage
\section{Introduction }


The Ising model holds a prominent position in  the history of physics and the development of various  physics fields, including condensed matter physics and statistical mechanics. Being simply an arrangement of  spin variables, with two possible values (representing molecules, particles, neurons, etc), and subjected to nearest-neighbour interactions with an additional local external field, it amazingly finds applications in a wealth of natural (from biological neural networks to ecology or the spread of diseases), artificial (e.g. artificial intelligence) and social systems (e.g. voters model). Its  simpler version in one dimension was solved initially by Ernst Ising, after the proposal of his supervisor, Wilhem Lenz, to solve it in higher dimensions, not knowing that the complexity it bore had to wait for many years of advancement of science to be solved~\cite{1967Brush,2017Ising}. 
Our interest in the Ising model in one dimension here is substantiated by its capability  to encode mathematical and computational problems. There is a famous correspondence of Ising models with quadratic unconstrained binary optimization problems (QUBO) with boolean variables. This lies the connection of finding the ground state of the Ising model to optimization problems, and therefore to the solution of problems of interest, such as the Boolean satisfiability problem~\cite{1984Tovey,2008Marques}. Indeed, the connection with satisfiability problems allowed to establish that finding the ground state is a NP-hard problem for two and three dimensions~\cite{1982Barahona}. 

The growth of the number of possible configurations as $2^N$, for $N$ spins is ultimately behind this complexity, and this points to a clear connection: what if spin variables are modelled as quantum two-state variables, i.e. qubits, and one allows the problem to perform a quantum evolution towards its ground state? this is the underlying idea with the utilization of quantum systems to simulate spin Ising models~\cite{1982Benioff}. This connection led to two strategies: to develope approximate methods, often quantum inspired, to solve the problem; and to use directly a quantum computer. 


The first strategy derived in a list of techniques, such as the Density Matrix Renormalization Group (DMRG) \cite{PhysRevLett.69.2863,RevModPhys.77.259},
Matrix Product States, and related methods (see, for example \cite{Cirac2009},\cite{PhysRevLett.93.040502}). 
One can also make use of methods which are based on the Lie-Trotter
formula \cite{PhysRevX.11.011020}, Krylov subspace expansion \cite{Saad1992},
truncated Taylor series  \cite{Truuncated_Taylor22} or randomized
product formulae \cite{PhysRevLett.123.070503,PRXQuantum.2.040305,Faehrmann2022randomizingmulti}. This collection of techniques allowed to obtain results fruitful also for the very development of physics.
We address the reader to \cite{Miessen2023} for a recent review regarding a comparative performance of these methods.


The second strategy required the development and reliable functioning at sufficiently large sizes of quantum computers. 
One option is to mimic the well known Monte Carlo-based method of simulated annealing, termed as quantum annealing. This has been implemented even in commercial set-ups for thousands of qubits~\cite{2010Berkley,2010Bian}. Also, one can use quantum gate-based computers and implement various algorithms, see e.g.~\cite{CerveraLierta2018exactisingmodel}.
Most of the above approximate methods can also be used for this purpose,
or in combination with hybrid optimization methods \cite{Mansuroglu2023,PhysRevLett.123.070503}. 

This second strategy, which evidently can lead to the most efficient methods, finds its limitation in the number of practical qubits available. For quantum annealing methods, current technology allows for the order of a few thousands, while for quantum gate systems it is at present limited to a few hundreds (see e.g.~\cite{ choi2023ibm}).
But the practical problems, e.g., boolean satisfiability, require  even millions of variables~\cite{2002Zhang}. 


In this technological context,  reducing the number of qubits is crucial for being able to simulate larger many-body systems on current available NISQ (Noisy Intermediate-Scale Quantum) processors.
In this paper, we introduce and test a tool which can be used to reduce the number of qubits needed to solve a particular problem by one half for Ising models. To this end,  we employ a well-established theoretical framework known
as Renormalization Group theory (RG), which has been developed over
the course of many years. The RG theory involves a series of systematic
steps that aim to reduce the number of spins in the system while simultaneously
redefining the original couplings between them. These steps are iterated
multiple times, leading to a progressive reduction in the number of
spins. By employing this iterative process, researchers
are able to extract valuable information regarding the critical behavior
of the system, contributing to our understanding of its fundamental
properties \cite{PhysRev.95.1300,PhysicsPhysiqueFizika.2.263,RevModPhys.47.773,PhysRevB.4.3184}. 


In particular, we concentrate on the Block  RG Method (BRGM), which
makes use of the concept of blocks already introduced by Kadanoff
\cite{PhysicsPhysiqueFizika.2.263,PhysRevD.16.1769,PhysRevB.18.3568,fernandez-pacheco_comment_1979}.
This method shares some similarities with the DMRG method. In both
cases, one makes a choice of such blocks and performs some truncation
on the eigenstates. However, while in the latter approach one chooses
the eigenstates of the corresponding density matrix with the
highest eigenvalues, for the BRGM one keeps the lowest energy states
of the block Hamiltonian, i.e., we perform the RG procedure in “real
space”. As it has been shown, a wise choice of these blocks and,
in particular, of the distribution of spin couplings inside and among
the blocks (the so-called “intrablock” and “interlock” terms,
respectively) can provide a good approximation to the study of critical
phenomena in these systems \cite{fernandez-pacheco_comment_1979,martn-delgado_renormalization_1996,PhysRevE.87.032154,Monthus2015}. 

The focus of our investigation implies a radically new approach, as it revolves around determining whether the BRGM iteration process yields meaningful results {\em within a finite number of steps}. This approach is totally different from the original scope of the BRGM method, which was intended to investigate critical points by performing an infinite iteration of such steps. To address this, we have conducted a thorough examination of the BRGM applied to the Ising model, utilizing a comparative analysis of relevant quantities. Our methodology involved systematically reducing the number of spins by a factor of 1/2 while appropriately adjusting the coupling constants. Remarkably, even with a relatively small number of spins, we have observed an excellent agreement between the obtained results, and the ones that correspond to the original larger Hamiltonian. 

An additional goal of this work is to show that the same method holds in generality and can be applied to other many-body Hamiltonians. To illustrate this, we have also applied the methodology to the one-dimensional anisotropic Heisenberg Hamiltonian. The Ising model is actually a simplified version of  the  Heisenberg model, also in one dimension,  where the interaction between spins occurs only in one direction, and a magnetic field is applied in a perpendicular direction.  The Heisenberg model was formulated to model magnetic materials, and has a prominent role in the study of criticality, phase transitions, entanglement, quantum field theories, and a long list.   For the Heisenberg model, BRGM method allows reducing the number of spins by a factor 1/3  \cite{martn-delgado_renormalization_1996}. Our numerical calculations allow us to show convergence of the results. But  in this case the computational requirements are such demanding, that a total resemblance between the results from the initial and renormalized Hamiltonians is not achieved. 


The findings with the Ising and more general Heisenberg  Hamiltonians highlight the efficacy of the BRGM iteration procedure in accurately characterizing the relevant properties of the many-body models, supporting its utility as a valuable tool in the study of complex physical systems using an effectively reduced number of spins. 

The structure of this paper is organized as follows: Sec.~\ref{sec:BRGM} offers
a brief overview of the theoretical foundations of the BRGM.
We delve into the key concepts and principles underlying this approach,
providing a concise review of the theory and its application in the
context of the Ising model. In Sec.~\ref{sec:BRGMHeisenberg} we apply these concepts to the Heisenberg model. Building upon this theoretical foundation,
Sec.~\ref{sec:physics}  focuses on our investigation of the physical features to be analyzed using the BRGM approach. We introduce the magnetization,
correlation functions, and entanglement entropy of a system with a given 
number of spins. Section~\ref{sec:results} is dedicated
to the presentation and discussion of our findings. We compare and
contrast the results for different spin numbers, and discuss their implications. In Sec.~\ref{sec:conv} we perform a detailed analysis that allows to visualize the convergence features we found within the BRGM. Finally, in Sec.~\ref{sec:conclusions} we provide a comprehensive conclusion based on our results, and discuss the broader implications of our study. We ofer further discussion on convergence in App.~\ref{app:T}. 

	\section{BRGM Approach to the Ising Model }
 \label{sec:BRGM}

Let us consider a Hilbert space \( \mathcal{H} \) with a Hamiltonian \( H \). As explained in \cite{martn-delgado_renormalization_1996}, the BRGM is defined by an embedding operator \( T : \mathcal{H'} \rightarrow \mathcal{H} \), and its corresponding truncation operator \( T^\dagger : \mathcal{H} \rightarrow \mathcal{H'} \), where \( \mathcal{H'} \) is a new Hilbert space where the renormalized Hamiltonian \( H' \) is defined.

\begin{figure}[ht]
  \centering
  \includegraphics[width=0.8\textwidth]{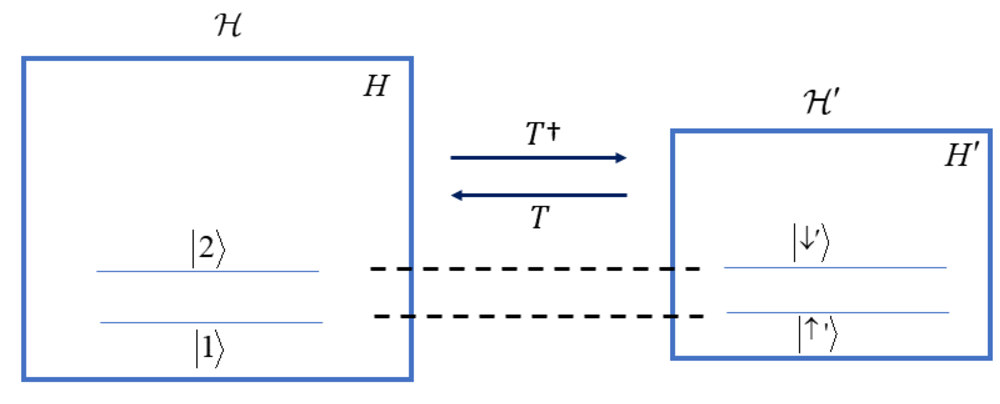}
  \caption{Schematic of the BRGM conceptual setting. The embedding operator is \( T : \mathcal{H'} \rightarrow \mathcal{H} \), and its corresponding truncation operator is  \( T^\dagger : \mathcal{H} \rightarrow \mathcal{H'} \), where  \( \mathcal{H}  \) is the original Hilbert space  and \( \mathcal{H'} \)  is  a new Hilbert space associated to the renormalized Hamiltonian \( H' \). }
  \label{Fig1new}
\end{figure}

The criterion to be accomplished is that \( H \) and \( H' \) have in common their low-lying spectrum (see Figure~~\ref{Fig1new}).
 \FloatBarrier
\FloatBarrier
\captionsetup{font=small} 

Thus, given \( |\psi'\rangle \) in \(\mathcal{H'}\), we define \( |\psi\rangle \) in \( \mathcal{H} \) such that
\begin{equation}
|\psi\rangle = T|\psi'\rangle,
\label{Eq.1}
\end{equation}
and we impose that \cite{martn-delgado_renormalization_1996}:
\begin{equation}
T^\dagger T = I_\mathcal{H'} ,
\label{Eq.2}
\end{equation}
with $I_\mathcal{H'}$ the identity operator in $\mathcal{H'}$, so that 
\begin{equation}
|\psi'\rangle = T^\dagger|\psi\rangle.
\label{Eq.3}
\end{equation}
These operators can be used to define the new Hamiltonian \( H' \) as
\begin{equation}
H' = T^\dagger H T.
\label{Eq.4}
\end{equation}
Notice that the inverse of Eq. (\ref{Eq.2}) does not hold in general, i.e., the operator
\begin{equation}
P \equiv T T^\dagger \neq I_\mathcal{H}, 
\label{Eq.5}
\end{equation}
where $I_\mathcal{H}$ is the identity operator in $\mathcal{H}$. Otherwise, such an equality, together with Eq. (\ref{Eq.2}), would imply that \( \mathcal{H} \) and \( \mathcal{H'} \) are isomorphic, while the truncation inherent to the renormalization method assumes that \( \dim \mathcal{H'} < \dim \mathcal{H} \).

The method we follow here is based on a block spin transformation that preserves the structure of the model \cite{PhysRevE.87.032154}. To illustrate it, let us  start from the Ising Hamiltonian subject to an external magnetic field,  defined as
 \begin{equation}
		H=-\sum_{i=1}^{N}J_{i}\sigma_{i}^{z}\sigma_{i+1}^{z}-\sum_{i=1}^{N}\Gamma_{i}\sigma_{i}^{x},
         \label{Eq.6}
	\end{equation}
where $\sigma_{i}^{z}$ and $\sigma_{i}^{x}$ are Pauli matrices acting on the $i$th spin, and $N$ is assumed to be even. We assume periodic boundary conditions. Then we divide the chain into blocks of two spins. We assume this Hamiltonian encodes in the coefficients $J_{i}$ and $\Gamma_{i}$ some practical problem one aims to solve. 
 
 \FloatBarrier
\begin{figure}[ht]
  \centering
  \includegraphics[width=0.8\textwidth]{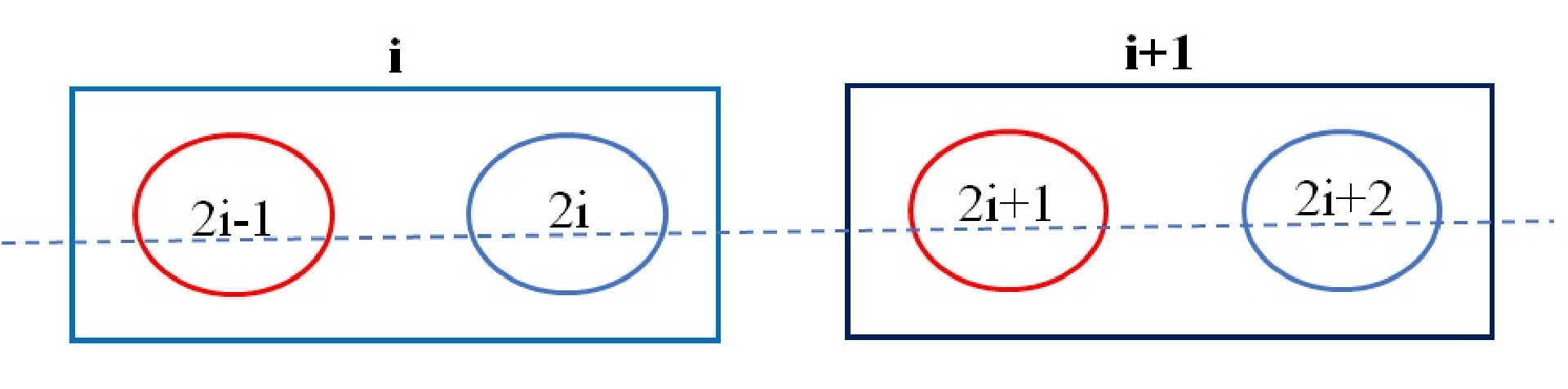}
  \caption{Generation of block spin formations in a one-dimensional setting.}
  \label{fig:Fig1}
\end{figure} 
\FloatBarrier
\captionsetup{font=small} 

  The initial Hamiltonian is split into intrablock and interblock parts~\cite{miyazaki2011real}  (see  Figure. \ref{fig:Fig1}) giving rise to
  \begin{equation}
  H_{i}^{\rm{intra}}=-J_{2i-1}\sigma_{2i-1}^{z}\sigma_{2i}^{z}-\Gamma_{2i-1}\sigma_{2i-1}^{x},
  \label{Eq.7}
  \end{equation}
 \begin{equation}
	H_{i,i+1}^{\rm{inter}}=-J_{2i}\sigma_{2i}^{z}\sigma_{2i+1}^{z}-\Gamma_{2i}\sigma_{2i}^{x},
 \label{Eq.8}
\end{equation}
where spins $2i\!-\!1$ and $2i$ belong to block $i$,  and spin $2i\!+\!1$ (together with $2i+2$) belongs to block $i\!+\!1$. The label $i$ of block then runs from 1 to N/2.
Now, we compute the eigenvalues of $H_{i}^{\rm{intra}}$ as follows
\begin{subequations}
\begin{equation}
\varepsilon_{i}^{(1)}=\varepsilon_{i}^{(2)}=-\sqrt{(J_{2i-1})^{2}+(\Gamma_{2i-1})^{2}},
\label{Eq.9a}
\end{equation}
\begin{equation}
\varepsilon_{i}^{(3)}=\varepsilon_{i}^{(4)}=\sqrt{(J_{2i-1})^{2}+(\Gamma_{2i-1})^{2}}.
\label{Eq.9b}
\end{equation}
\end{subequations}
As we see, these are degenerate.
The corresponding eigenvectors are
\begin{subequations}
\begin{equation}
    	\ket{1}_{i}=a_{i}^{+}\ket{\uparrow\uparrow}+a_{i}^{-}\ket{\downarrow\uparrow},
 \label{Eq.10a}
\end{equation}
\begin{equation}
	\ket{2}_{i}=a_{i}^{+}\ket{\downarrow\downarrow}+a_{i}^{-}\ket{\uparrow\downarrow},
 \label{Eq.10b}
\end{equation}
\begin{equation}
	\ket{3}_{i}=a_{i}^{-}\ket{\downarrow\downarrow}-a_{i}^{+}\ket{\uparrow\downarrow},
 \label{Eq.10c}
 \end{equation}
\begin{equation}
	\ket{4}_{i}=a_{i}^{-}\ket{\uparrow\uparrow}-a_{i}^{+}\ket{\downarrow\uparrow},	
 \label{Eq.10d}
  \end{equation}
\end{subequations}
where
\begin{equation}
	a_{i}^{\pm}=\sqrt{\frac{1}{2}(1\pm\frac{J_{2i-1}}{(J_{2i-1})^{2}+(\Gamma_{2i-1})^{2}})},
 \label{Eq.11}
\end{equation}
and $\ket{\uparrow\uparrow},\ket{\downarrow\uparrow},\ket{\uparrow\downarrow},\ket{\downarrow\downarrow}$ is the orthonormal basis in the 
$\sigma_{z}$ basis, i.e, $\sigma_{z}\ket{\uparrow}=\ket{\uparrow},$ $\sigma_{z}\ket{\downarrow}=-\ket{\downarrow}$. 
We keep the two lowest-lying energy eigenstates $\ket{1}_i$ and $\ket{2}_i$,  and drop the others. We then replace each block with a single spin represented by the $\ket{\uparrow '}$ and $\ket{\downarrow '}$ states. 
To this end, we define the embedding operator as
\begin{equation}
	T=\bigotimes_{i=1}^{N/2}T_{i},
 \label{Eq.12}
\end{equation}
with
\begin{equation}
	T_{i}=(\ket{1}\bra{\uparrow '}+\ket{2}\bra{\downarrow '})_{i}.
 \label{Eq.13}
\end{equation}
The resulting coarse-grained Hamiltonian $H_{\rm{RG}}$ is defined by the projection 
\begin{equation}
H_{\rm{RG}} \equiv H'=T^\dagger H T=\sum_{i=1}^{N/2}\varepsilon_{i}^{(1)}1_{i}-\sum_{i=1}^{N/2}\tilde{J_{i}}\tilde{\sigma_{i}^{z}}\tilde{\sigma_{i+1}^{z}}-\sum_{i=1}^{N/2}\tilde{\Gamma_{i}}\tilde{\sigma_{i}^{x}},
\label{Eq.14}
	\end{equation}
where the renormalized couplings are
\begin{subequations}
    \begin{equation}
	\tilde{J_{i}}=\frac{J_{2i}J_{2i+1}}{\sqrt{(J_{2i+1})^{2}+(\Gamma_{2i+1})^{2}}},
 \label{Eq.15a}
\end{equation}
\begin{equation}
	\tilde{\Gamma_{i}}=\frac{\Gamma_{2i-1}\Gamma_{2i}}{\sqrt{(J_{2i-1})^{2}+(\Gamma_{2i-1})^{2}}} .
 \label{Eq.15b}
\end{equation}
\end{subequations}
As we can see, this transformation preserves the form of the initial Hamiltonian defined in Eq.~\eqref{Eq.6} except for the first term in Eq.~\eqref{Eq.14}, which is proportional to the identity \cite{PhysRevE.87.032154}. 

\section{BRGM Approach to the Heisenberg Model}
 \label{sec:BRGMHeisenberg}
Let us now illustrate the method with a different  1d-lattice Hamiltonian, the Heisenberg	model. The Hamiltonian for this model is given by \cite{martn-delgado_renormalization_1996}
	\begin{equation}
		H_N = J \sum_{j=1}^{N-1} \left( S_j^x S_{j+1}^x + S_j^y S_{j+1}^y + \Delta S_j^z S_{j+1}^z \right),
  \label{Eq.16}
	\end{equation}
with $\Delta$  an  anisotropic parameter which is constrained to values greater than or equal to zero, while $J$ must be positive for the antiferromagnetic scenario \cite{white1983quantum}. When $\Delta$ equals 1, the system conforms to the antiferromagnetic-Heisenberg model, famously solved by Bethe in 1931 \cite{bethe1931theory}. Conversely, when $\Delta$ is set to 0, the system resembles the XY-model \cite{lieb1961two}. An interesting aspect of this scenario is the ease with which it can be solved using a Jordan-Wigner transformation, effectively mapping it to a free fermion model \cite{jordan1993paulische}. For other values of $\Delta$, Bethe's ansatz remains a viable method for solution, revealing its nature as a one-dimensional analogue to the 2D statistical mechanical model known as the 6-vertex or XXZ-model \cite{baxter2007exactly}.
		
		Regarding the RG-approach to half-integer spin or fermion models, it is common practice to focus on blocks comprising an odd number of sites. This choice, while not mandatory, often leads to effective Hamiltonians mirroring the original ones. By employing blocks consisting of three sites and conducting subsequent computations \cite{martn-delgado_renormalization_1996}, one arrives at the Renormalized Hamiltonian 
		\begin{equation}
			H_{\rm{RG}} = \frac{N}{3} E_{B}(J, \Delta) + H_{N/3}(J', \Delta').
        \label{Eq.17}
		\end{equation}
		This Hamiltonian encapsulates the essence of our reduced system, where the number of spins has been decreased by a factor of 1/3. 
		It comprises two main components: a term related to the magnetic field energy $E_{B}(J, \Delta)$, and a renormalized Hamiltonian $H_{N/3}(J', \Delta')$ that characterizes the interactions among the reduced spins, where the parameters $ J' = (\xi^x)^2 J $ and $\Delta' = \left( \frac{\xi^z}{\xi^x} \right)^2 \Delta$, crucial for determining the behavior of our system, are defined in terms of $\xi^x$ and $\xi^z$ respectively as 
		
			\begin{equation}
			\xi^x = \xi^y \equiv \frac{2(1 + x)(1 - 2x)}{3(1 + 2x^2)},
        \label{Eq.19}
		\end{equation}
  and	
		\begin{equation}
			\xi^z \equiv \frac{2(1 + x)^2}{3(1 + 2x^2)}, 
        \label{Eq.20}
		\end{equation}

where

   \begin{equation}
		 	x = \frac{2(\Delta - 1)}{8 + \Delta + 3\sqrt{\Delta^2 + 8}}.
        \label{Eq.18}
		 \end{equation}
    To capture the contribution of the magnetic field to the overall energy of the system we introduce  the energy term $E_B$, which has the expression
	   	\begin{equation}
	   	E_B = -\frac{J}{4} \left[ \Delta + \sqrt{\Delta^2 + 8} \right].
         \label{Eq.21}
	   \end{equation}
	   These definitions provide a solid foundation for our subsequent analysis of the physical properties of the re-normalized system. 
		
\section{Investigation of Physical Features with the Use of the BRGM}
\label{sec:physics}

In what follows, we will investigate the application of the BRGM  approach to an Ising and Heisenberg spin chain system with different sizes, i.e., to different values of $N$. Our aim is to apply the renormalization procedure to each chain, which implies a reduction in the number of spins by a factor of $1/2$ and $1/3$ for the Ising and Heisenberg model, respectively. We will examine key observables, including magnetization, correlation functions, and entanglement entropy, which will be compared to calculations obtained with the original Hamiltonian. We will progressively increase the number of initial spins, by considering the cases with $N=6$, $N=12$, and $N=24$ for the Ising model, and then $N=12$ and $N=24$ for the Heisenberg model. 


In all cases, we start from some initial state $\ket{\psi_0}$, and perform the time evolution with the initial Hamiltonian $H$. The evolved state is then used to calculate the magnitudes mentioned above, which are compared to the ones derived from the corresponding initial state 
\begin{equation}
	   	\ket{\psi'_0}=T^\dagger \ket{\psi_0},
         \label{Eq.21b}
	   \end{equation}
whose time evolution takes place according to the renormalized Hamiltonian $H_{RG}$.
Analogously to Eq.~(\ref{Eq.4}), an observable $A$ defined in the Hilbert space $\mathcal{H}$ is mapped to its corresponding observable $A'$ in $\mathcal{H'}$ using 
\begin{equation}
	   	A'=T^\dagger A T.
         \label{Eq.21c}
	   \end{equation}

\subsection{Magnetization}
 Magnetization is a property of magnetic materials that describes the degree of alignment of its spins. To compute the magnetization, one computes the expected value of the individual spins on the present state, and performs a sum over them. The formula to compute the average magnetization $M(t)$  at time $t$ for the considered system with $N$ spins is therefore \cite{islam2013emergence}
 \begin{equation}
 M(t)=\frac{1}{N}\sum_{i=1}^{N}\bra{\Psi(t)}\sigma_{z}^{i}\ket{\Psi(t)},
 \label{Eq.22}
 \end{equation}
 where $\ket{\Psi(t)}$ denotes the quantum state of the system at time $t$, and the summation is taken over all $N$ spins in the system.  The average magnetization can be positive, negative, or zero depending on the orientation of the individual magnetic moments and the strength of the interactions between them.

\subsection{Spin Correlation Functions}
A spin correlation function is a quantity that characterizes the degree of correlation between the spins at different positions. The spin correlation function at  time $t$  is defined as 
\begin{equation}
	C^{i}_r(t)=\bra{\Psi(t)} \sigma_{z}^{i} \sigma_{z}^{i+r}\ket{\Psi(t)}-\bra{\Psi(t)}\sigma_{z}^{i}\ket{\Psi(t)} \bra{\Psi(t)} \sigma_{z}^{i+r}\ket{\Psi(t)},
 \label{Eq.23}
\end{equation}
where we are comparing spins at $i$ and $i\!+\!r$ positions, and we use periodic boundary conditions. 

\subsection{Entanglement Entropy}
Entanglement entropy is a quantity that characterizes the amount of quantum entanglement between two parts of a larger quantum system. The entanglement entropy is computed by dividing the system into two parts, $A$ and $B$, and then tracing out the degrees of freedom in the region $B$ to obtain the reduced density matrix for region $A$~\cite{amico2008entanglement, calabrese2004entanglement}. In this work, the entanglement entropy $S$ is computed as the Von Neumann entropy of this reduced density matrix, as follows 
\begin{equation}
	S=-\rm{Tr}(\rho_{A} log\rho_{A}),
 \label{Eq.24}
\end{equation}
where $\rho_{A}$ is the reduced density matrix for region $A$, obtained by tracing out the degrees of freedom in region $B$, and $\rm{Tr}$ denotes the trace over the Hilbert space of the region $A$. In our analysis, we consider two regions, $A$ and $B$, with $x = 1$ spins in region $A$ and $N - x = N - 1$ spins in region $B$. It is worth noting that for the special case when $x = N/2$, the result remains unchanged. In this scenario, the entanglement entropy still exhibits a logarithmic scaling with the size of the boundary between regions $A$ and $B$, indicating the presence of long-range entanglement in the system. The entanglement entropy can be used to study the behavior of the system as a function of the magnetic field strength, and to identify phase transitions and critical points in the system~\cite{eisert2010colloquium}.

 \section{Results and Discussion}
\label{sec:results}

 We have developed a code with Qutip (an open-source software for simulating the dynamics of open quantum systems, see~\cite{johansson2012qutip,johansson2012qutipb}) where each spin is a qubit, and we performed the quantum evolution of the reduced RG Hamiltonian $H_{\rm{RG}}$ and the initial one, to compare among them. To perform such calculations,  we made use of the Lluis Vives supercomputer from University of Valencia, Spain. Lluis Vives is an Altix UltraViolet 1000 server from the Silicon Graphics company. It has 64 Xeon 7500 series hexacore CPUs at 2.67 GHz and 18 MB of on-die L3 cache, 2048 GB of RAM and about 15 TB of hard disk. 

\subsection{Ising Model}

We have calculated the above-defined magnitudes with an increasing even number of spins, finding a good agreement for $N=24$ spins, which is also already close to the limit we found of the computational capability at our disposal. These are proof-of-concept calculations. With these results, we show that one can  implement the reduced Hamiltonian in a real quantum computer and perform the evolution on the physical system to simulate with the RG Hamiltonian  $H_{\rm{RG}}$ given in Eq.~\eqref{Eq.14}. The results obtained will then be helpful to solve the practical problem which may be encoded in the coefficients of the corresponding  Hamiltonian of interest Eq.~\eqref{Eq.6}.  
 
In the following figures,  we depict two  different scenarios for each of the magnitudes defined in the previous Section, as one increases the number $N$ of spins. We show results for the original Hamiltonian $H$ Eq.(~\eqref{Eq.6}), and for the RG Hamiltonian $H_{\rm{RG}}$ with modified coefficients given in Eq.~\eqref{Eq.15a} and Eq.~\eqref{Eq.15b}. 
For the initial comparisons, we consider a homogeneous Ising model, for which all $J_i$ take the same value (indicated by $J$), and similarly for $\Gamma_i$, which will be denoted as $\Gamma$. At the end of this section we also analyze a more general situation, which involves assigning random values of the couplings at each site, to check the validity of our results. 

\begin{figure}[ht]
    \centering
    \includegraphics[width=1.00\textwidth]{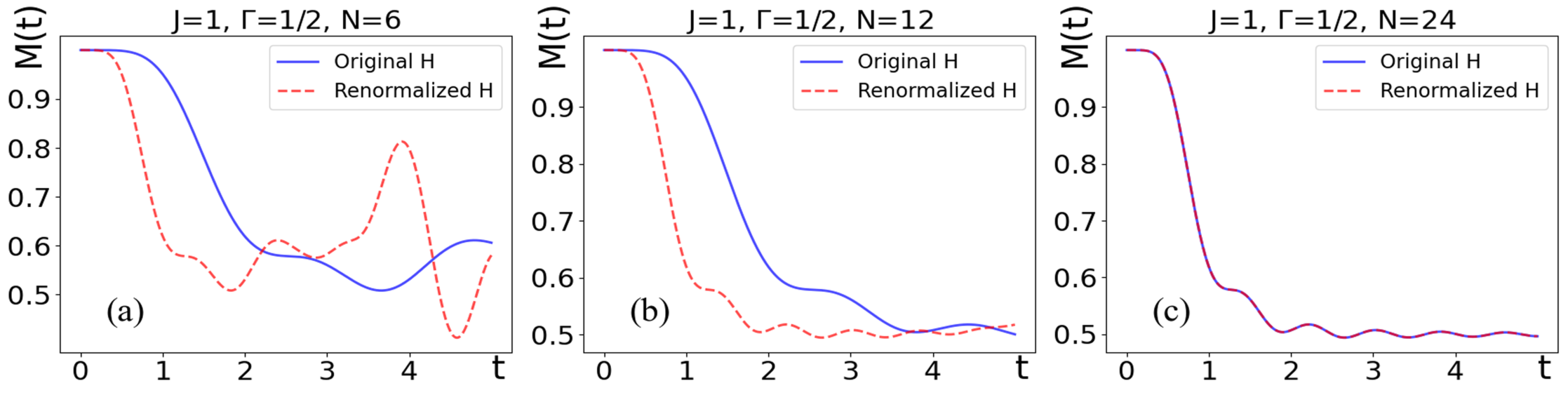}
    \caption{Total magnetization as a function of time. Panels  (a), (b), and (c) correspond to $N=6, 12$ and 24 spins, respectively. In each panel, the blue curve corresponds to the original Hamiltonian, while the red  curve is the result for the renormalized Hamiltonian $H_{\rm{RG}}$. The original coefficients are taken as $J=1$ and $\Gamma=1/2$. All magnitudes are adimensional. These plots correspond to the initial state where all spins are up.}
    \label{Fig3}
\end{figure}

 \autoref{Fig3} shows these results for the magnetization for different numbers of spins. Upon analyzing the figures representing the magnetization of the original Hamiltonian of the Ising model and its corresponding renormalized Hamiltonian as $N$ is increased, one observes a clear convergence between both of them.  Indeed, for the case of 24 spins, \autoref{Fig3} (c), the magnetization curves for both systems exhibit a remarkable coincidence. 
\FloatBarrier
\FloatBarrier
\captionsetup{font=small}

\begin{figure}[ht]
    \centering
    \includegraphics[width=0.35 \textwidth]{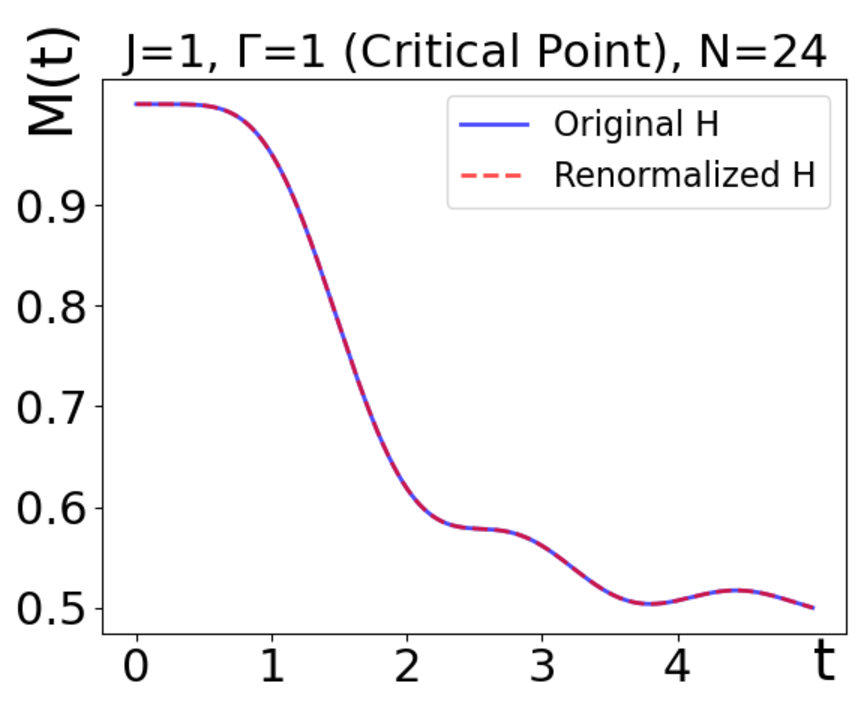}
    \caption{Total Magnetization vs. time at the critical point  \( J=\Gamma = 1\) for a system of 24 spins with an initial state where all spins are up.}
    \label{Fig4}
\end{figure}

In this study, we also examine the magnetization behavior at the critical point  \( J= \Gamma = 1\), specifically, for a system of 24 spins. Our results demonstrate that, even at this critical point, there is convergence between the magnetization profiles of the original and renormalized Hamiltonians. This observation suggests that criticality does not significantly affect the convergence between these two Hamiltonians. The corresponding results are presented in Figure~\ref{Fig4}.
\FloatBarrier
\FloatBarrier
\captionsetup{font=small}

\begin{figure}[ht]
    \centering
    \includegraphics[width=1.00\textwidth]{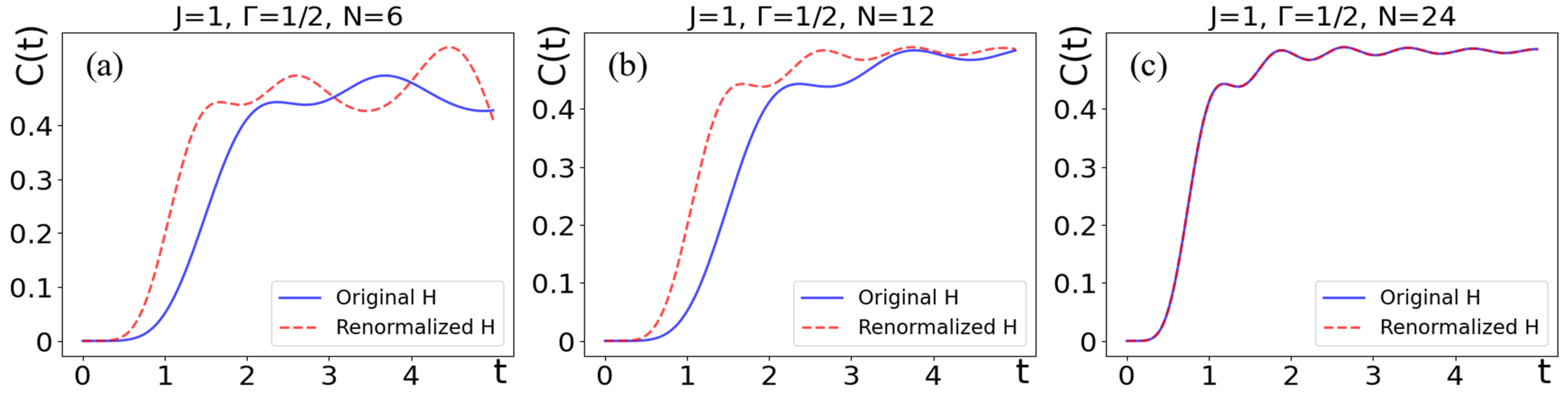}
    \caption{The spin correlation function given in Eq.~\eqref{Eq.23} for $i=1$ and $r=1$ as a function of time for (a) $N=6$ spins, (b) $N=12$ spins, (c) $N=24$ spins. The original coefficients are taken as $J=1$ and $\Gamma=1/2$. All magnitudes are adimensional. These plots correspond to the initial state where all spins are up.}
    \label{Fig5}
\end{figure}

 \autoref{Fig5} presents a comprehensive analysis of the spin correlation function as defined by Eq.~\eqref{Eq.23} for $i=1$ and $r=1$. Similar results are obtained for other values of $r$. We see again a clear convergence from the results derived by the BRGM Hamiltonian, with one half the original spins, towards the initial Hamiltonian. The degree of achieved coincidence for $N=24$ spins is striking. 
\FloatBarrier
\FloatBarrier
\captionsetup{font=small}    
   
\begin{figure}[ht]
    \centering
    \includegraphics[width=1.00\textwidth]{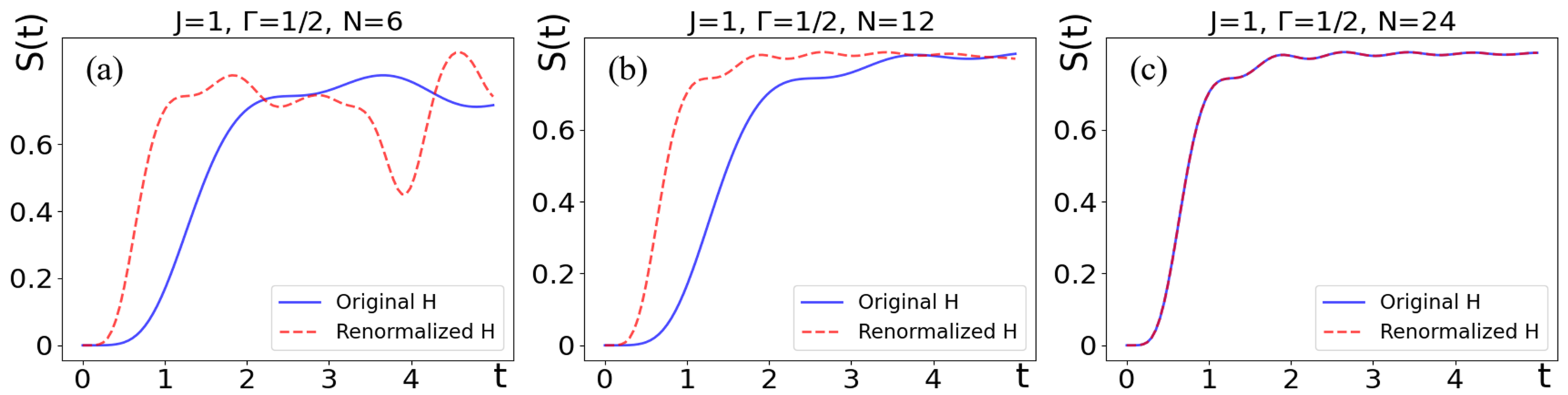}
    \caption{The entanglement entropy given in Eq.~\eqref{Eq.24} as a function of time for (a) $N=6$ spins, (b) $N=12$ spins, (c) $N=24$ spins.  The original coefficients are taken as $J=1$ and $\Gamma=1/2$. All magnitudes are adimensional. These plots correspond to the initial state where all spins are up.}
    \label{Fig6}
\end{figure}

\begin{figure}[ht]
    \centering
    \includegraphics[width=1.00\textwidth]{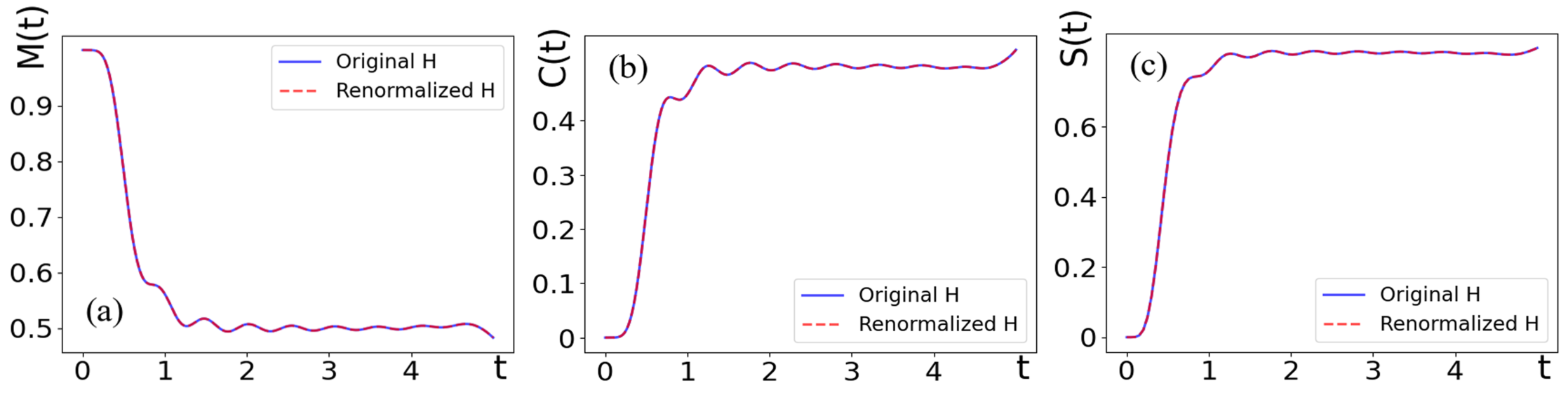}
    \caption{(a) Magnetization, (b) Spin correlation function, (c) Entanglement entropy as a function of time for 24 spins. The original coefficients $J_{i}$ and $\Gamma_{i}$ were obtained as random numbers within the interval [0,1]. All magnitudes are adimensional. These plots correspond to the initial state where all spins are up.}
    \label{Fig7}
\end{figure}

\begin{figure}[ht]
    \centering
    \includegraphics[width=1.00\textwidth]{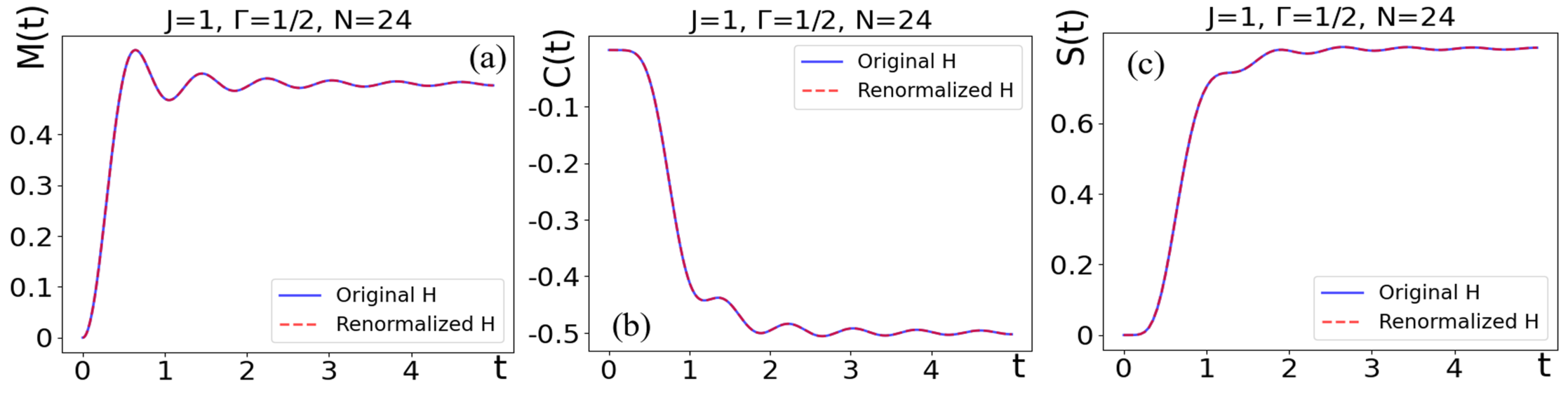}
    \caption{Comparison of the physical features for the initial state \( \ket{010101\ldots} \) with \( N = 24 \) spins: (a) Magnetization, (b) Spin correlation function, (c) Entanglement entropy.}
    \label{Initial}
\end{figure}

\autoref{Fig6} shows the entanglement entropy. The same comments hold in this case. A complete agreement is found for the case with $24$ spins.
\FloatBarrier
\FloatBarrier
\captionsetup{font=small}    
Moreover, we made calculations for an initial Hamiltonian with random coefficients, where $J_i$ and $\Gamma_i$ are randomly chosen  in the range [0,1]. We find the same results as in previous cases. We show directly the results for all three magnitudes and for $N=24$ in  Figure.~\ref{Fig7}. There is again a great overlap between the physical features of the original Hamiltonian and the renormalized Hamiltonian. We find that, regardless of the values for coefficients, the resulting outcome for physical features of a system containing 24 spins under BRGM is identical. 
\FloatBarrier
\FloatBarrier
\captionsetup{font=small}    
To conclude, we considered another initial state for our computation, specifically the alternating spin configuration \( \ket{010101\ldots} \). This configuration corresponds to a state where odd spins are up, and even spins are down. We then computed the three physical features for this system with \( N = 24 \) spins. The results are presented in Figure~\ref{Initial}. Remarkably, the findings indicate that for this initial state, we also observe convergence between the original Hamiltonian and the renormalized Hamiltonian, similar to our previous results. This illustrates that changing the initial states does not affect the overall outcome of our analysis. This conclusion is further supported by the argument we make at the end of Sec. 6, which is based on an operator analysis and is, therefore, state independent.

\FloatBarrier
\FloatBarrier
\captionsetup{font=small}

\subsection{Heisenberg Model}
\begin{figure}[ht]
    \centering
    \includegraphics[height=15cm]{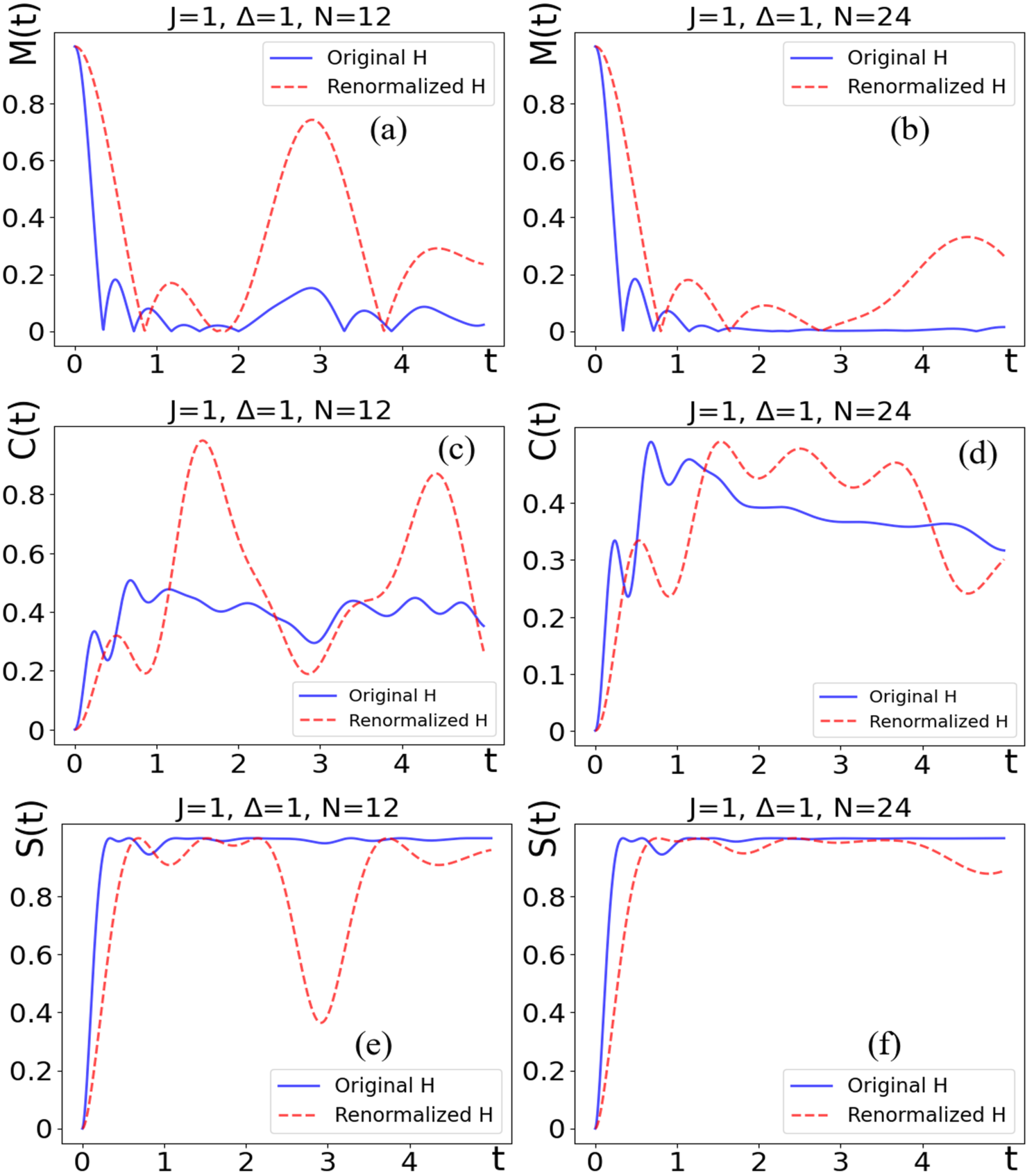}
\caption{Magnetization (a)--(b),  Spin correlation function (c)--(d), and Entanglement entropy (e)--(f), as a function of time for 12 and 24 spins, respectively. All magnitudes are adimensional. These plots correspond to the initial state where all spins are up.}
    \label{Fig8}
\end{figure}

	In this stage, we delve into examining various physical properties such as magnetization, correlation functions, and entanglement entropy for both the original and re-normalized Heisenberg Hamiltonian.
	Our objective is to implement the renormalization procedure on each chain, leading to a reduction in the number of spins by a factor of 1/3. Similarly to our previous approach with the Ising model, we consider scenarios with an  increasing number of spins, aiming at a system with N=24 spins. In Figure~\ref{Fig8}, we display plots for two scenarios: $N=12$ and $N=24$, both with $J=\Delta=1$. 

 As seen in Figures~\ref{Fig8} (a)-(f), it appears that by increasing the number of spins, the results from the two Hamiltonians (original and renormalized) become closer (a similar behavior  to that found for  the Ising model). However, considering that in this case the number of spins has been reduced by a factor $3$ (as compared to $2$ for the Ising model), we expect coincidence of results to appear for a larger $N$. This limit was   not accessible for the computations in the Luis Vives supercomputer, since saturation in memory manifested before. We will quantify the achieved degree of convergence in the next section, both for the Ising and Heisenberg Hamiltonians.

\FloatBarrier
\FloatBarrier
\captionsetup{font=small}   

\begin{figure}[ht]
    \centering
    \includegraphics[height=15cm]{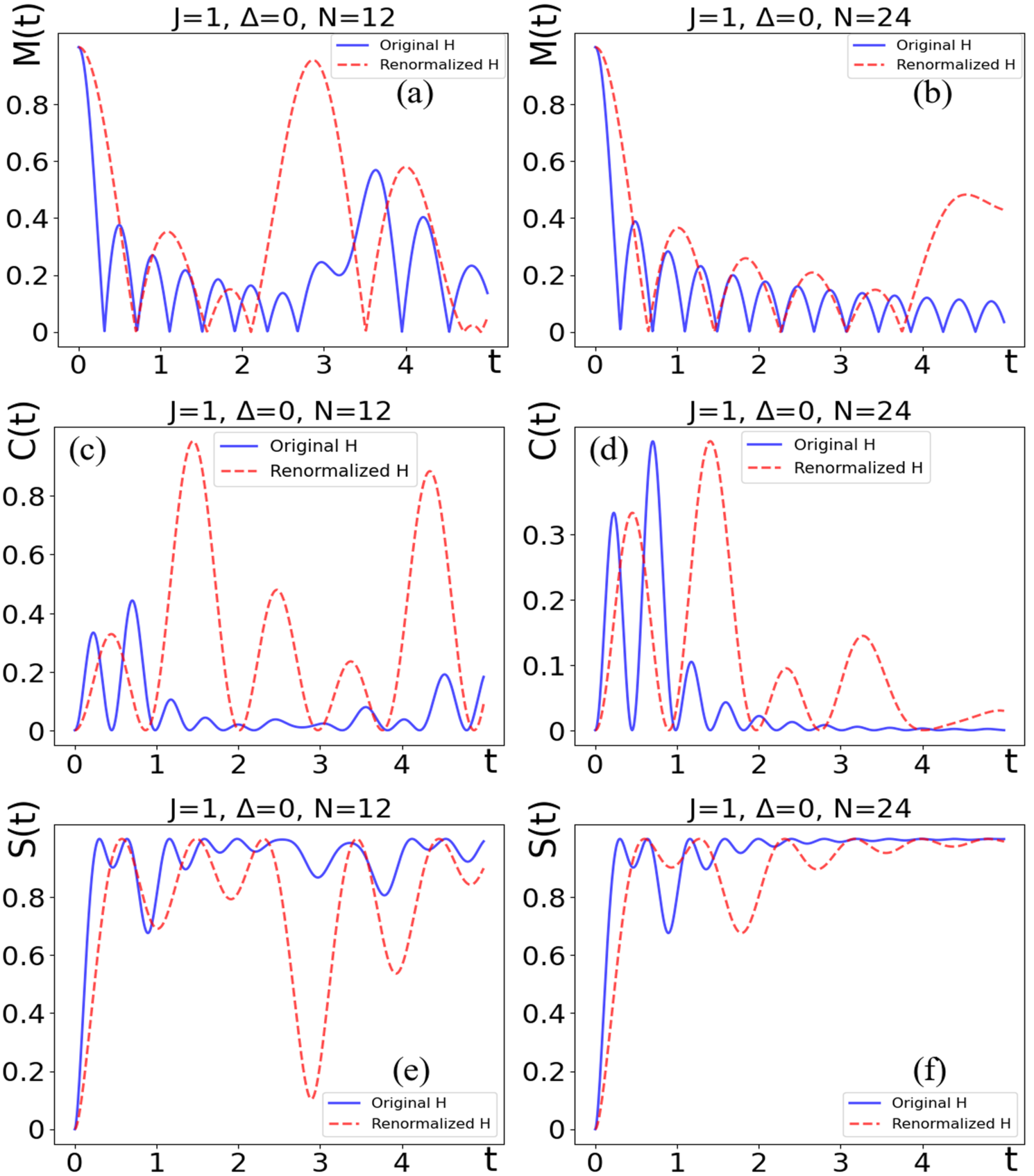}
\caption{Same as Fig.~\ref{Fig8}, with parameters $J=1$, $\Delta=0$.}
    \label{Fig9}
\end{figure}

\begin{figure}[ht]
    \centering
    \includegraphics[height=15cm]{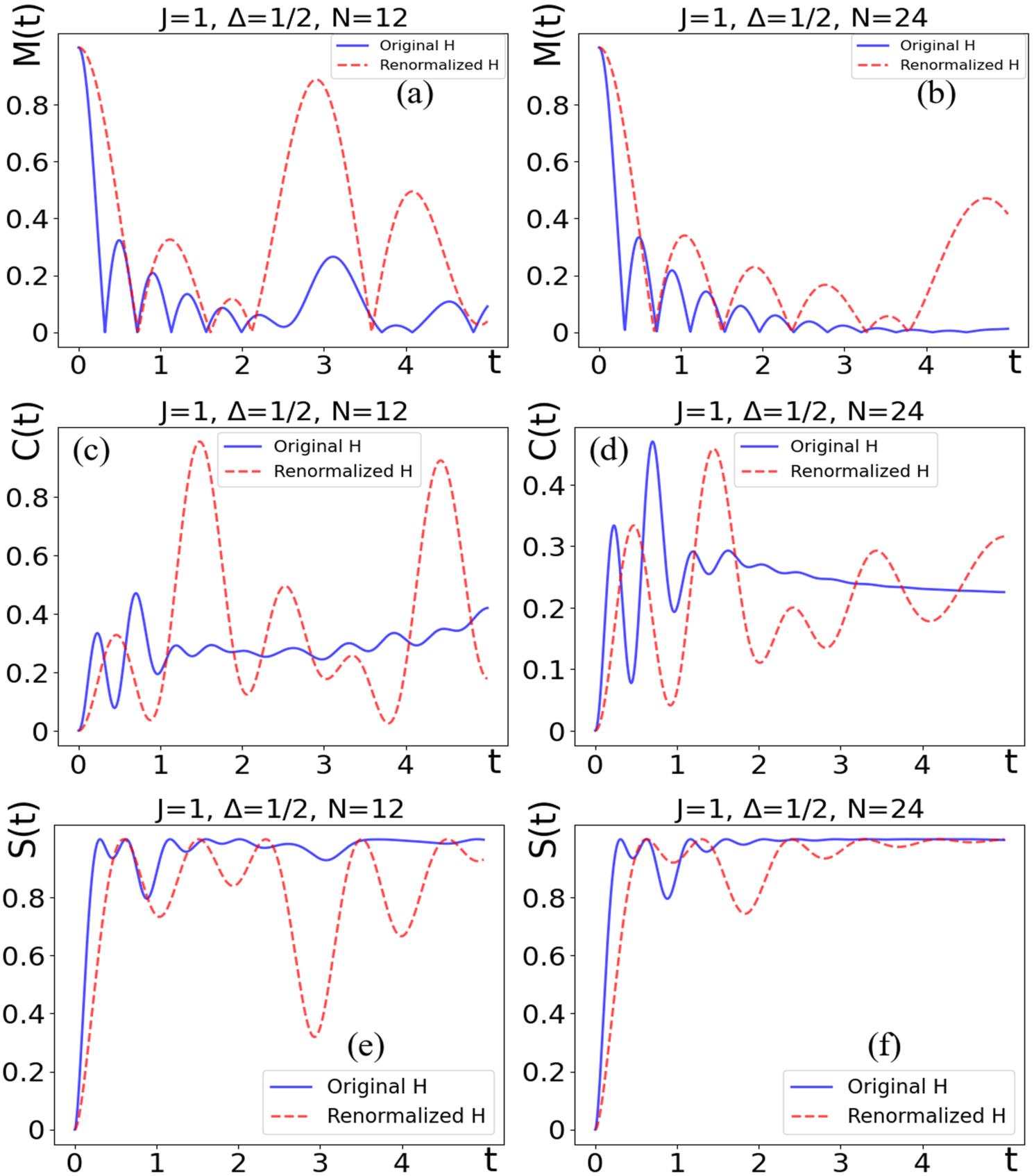}
\caption{Same as Fig.~\ref{Fig8}, with parameters $J=1$, $\Delta=1/2$.}
    \label{Fig10}
\end{figure}

We have also performed computations for the scenario where $J=1$ and $\Delta=0$. The results from these computations are shown in Fig.~\ref{Fig9}. Next, we extend our analysis to another set of parameters, specifically $J = 1$ and $\Delta = \frac{1}{2}$. The results of these computations are presented in Figure~\ref{Fig10}.

\FloatBarrier
\FloatBarrier
\captionsetup{font=small}

\FloatBarrier
\FloatBarrier
\captionsetup{font=small} 
   
\section {Convergence features for Ising and Heisenberg Hamiltonians}
\label{sec:conv}
The objective of this section is to quantify the degree of convergence which is achieved after the renormalization process, by comparing the curves obtained by both Hamiltonians for the different magnitudes we have been discussing i.e., magnetization, spin correlation function and entanglement entropy. More precisely, we will compute the chi-squared ($\chi^2$) for the difference of the two curves for a given magnitude, as a function of $N$, to investigate the convergence of the BRGM approximation as the number of spins is increased, for these two models.

\subsection{Ising Model}
Here, we compute the $\chi^2$ values for magnetization, entanglement entropy, and correlation function. We present three graphs corresponding to these physical features, with coefficients $J = 1$ and $\Gamma = 1/2$. By fitting the resulting plots to the function $a 2^{-bN}$, where $a$ and $b$ are constants, we demonstrate that the distance between the original and renormalized models decreases exponentially as the number of spins increases.
\FloatBarrier
\begin{figure}[ht]
    \centering
    \includegraphics[width=1.00\textwidth]{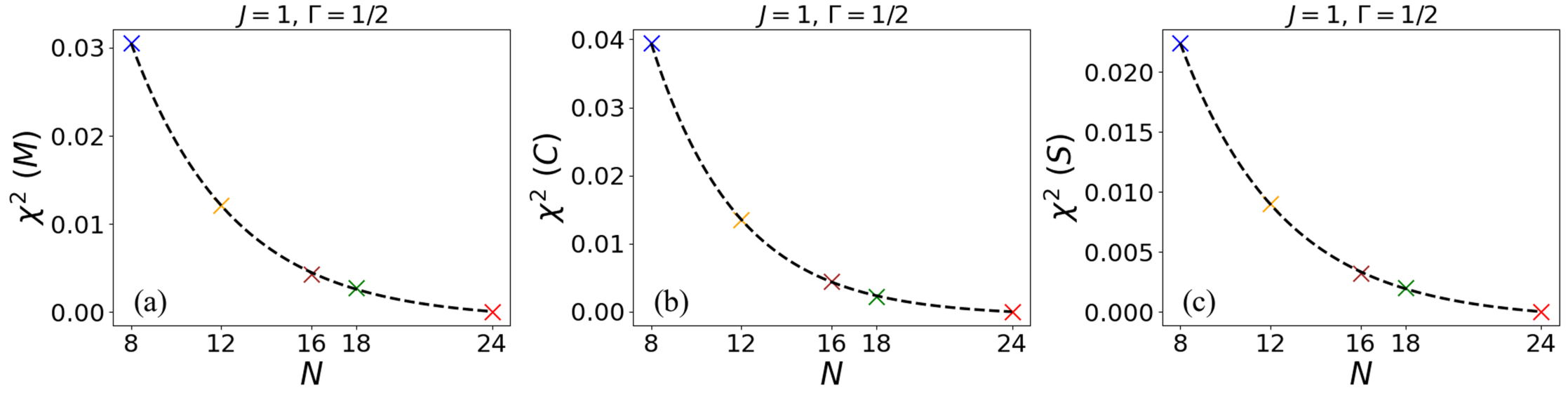}
\caption{Chi-squared value for the Ising model as a function of $N$ for (a) Magnetization, (b) Entanglement entropy, (c) Spin correlation function. These plots correspond to the initial state where all spins are up.}
    \label{Fig11}
\end{figure}
\FloatBarrier
\captionsetup{font=small} 
    
As shown in Fig.~\ref{Fig11}, the chi-squared  for all three physical features decreases with increasing number of spins, confirming that the distance between the two models diminishes exponentially. Additionally, for 24 spins, the $\chi^2$ value is very close to zero, indicating that the features of the original and renormalized Hamiltonians are equivalent at this system size, validating the effectiveness of the renormalization process for larger spin systems. The
detailed fitted parameters related to Figures. 12 (a)-(c) are listed in Table 1. 
\begin{table}[ht]
  \centering
  \begin{tabular}{|c|c|c|}
    \hline
    Figure & \(a\) & \(b\)  \\
    \hline
    Fig. 12(a) & 0.18 & 0.31 \\
    Fig. 12(b) & 0.32 & 0.37  \\
    Fig. 12(c) & 0.13 & 0.31  \\
    \hline
  \end{tabular}
  \caption{Fitted parameters \(a\), \(b\) for each part of Fig. 12}
  \label{table:Ising_fitted_parameters}
\end{table}

\subsection{Heisenberg Model}
\begin{figure}[ht]
    \centering
    \includegraphics[width=1.00\textwidth]{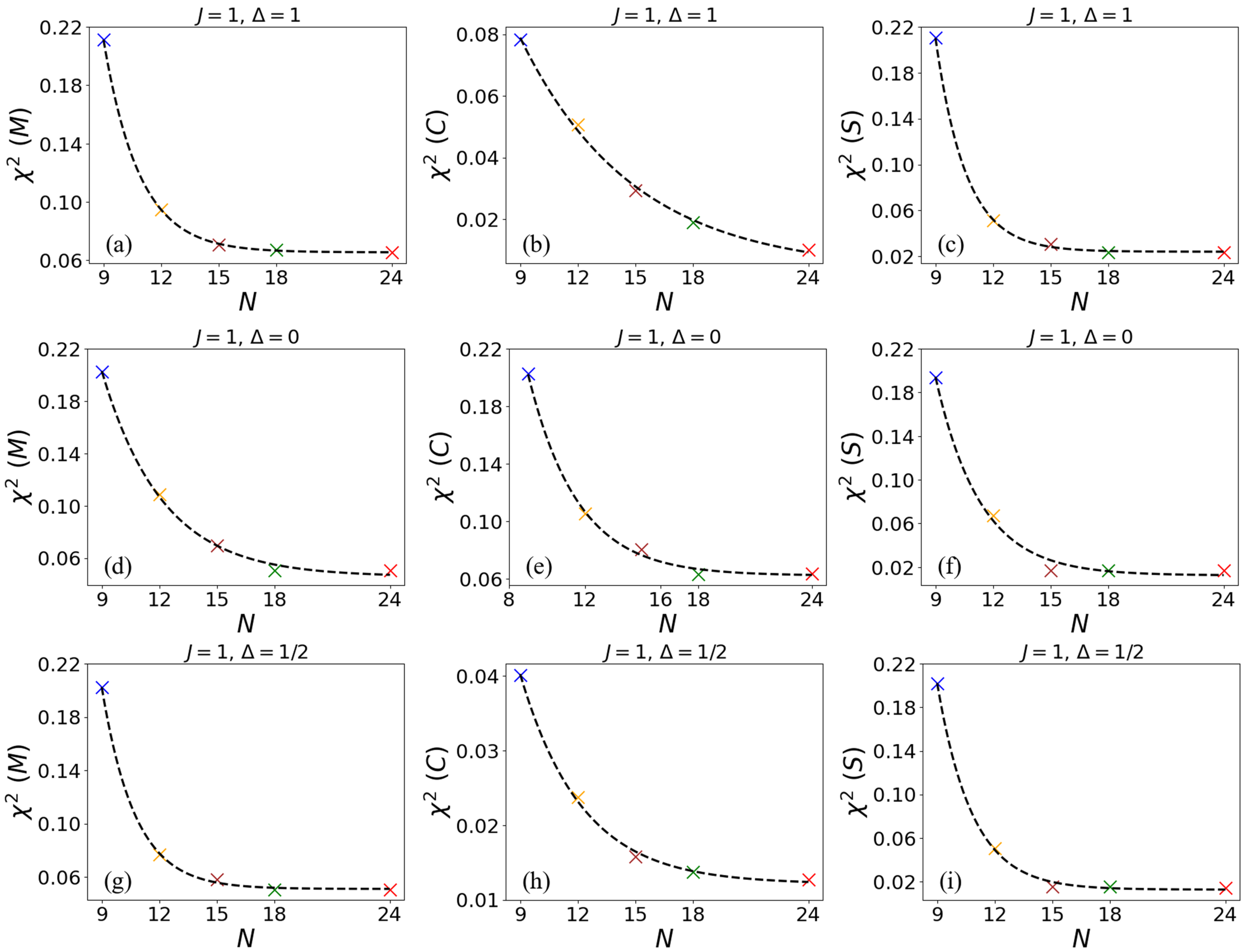}
\caption{Same as Fig.~\ref{Fig11}, now for the Heisenberg model, for three sets of parameters. These plots correspond to the initial state where all spins are up.}
    \label{Fig12}
\end{figure}

Now, we compute the $\chi^2$  for magnetization, entanglement entropy, and correlation function for three sets of parameters: $J = 1, \Delta = 1$, $J = 1, \Delta = 0$ and $J = 1, \Delta = 1/2$. We provide the results in Fig.~\ref{Fig12},  showing how the chi-squared  changes with the number of spins. As we discuss below, convergence is not completely achieved for this model, therefore plots will be fitted  to the function $a 2^{-bN}+c$, where the extra constant $c$ accounts for this incomplete convergence.
\FloatBarrier
\FloatBarrier
\captionsetup{font=small} 
    
As depicted in Fig.~\ref{Fig12}, the chi-squared  for the three sets of parameters exhibits an exponential decrease with increasing number of spins, up to an additive constant. The detailed fitted parameters related to Fig.~\ref{Fig12} are listed in Table~\ref{table:fitted_parameters}. We observe that, for some magnitudes (e.g., Fig. 13(b)), the value of $c$ is very small, and one sees a decrease which is close to an exponential. In other cases, as in Fig. 13(a), the achieved reduction is smaller, with a corresponding larger value of this coefficient. The question then is whether, by further increasing $N$, one would reach a complete convergence (or at least a much better improvement). Unfortunately, this is  beyond the capabilities of the supercomputer used for these calculations.  
\begin{table}[ht]
  \centering
  \begin{tabular}{|c|c|c|c|}
    \hline
    Figure & \(a\) & \(b\) & \(c\)\\
    \hline
    Fig. 13(a) & 18.09 & 0.77 & 0.060 \\
    Fig. 13(b) & 0.34 & 0.24 & 0.003\\
    Fig. 13(c) & 54.75 & 0.91 & 0.024\\
    Fig. 13(d) & 2.69 & 0.45 & 0.045\\
    Fig. 13(e) & 4.33 & 0.55 & 0.062\\
    Fig. 13(f) & 8.64 & 0.61 & 0.012  \\
    Fig. 13(g) & 27.44 & 0.83 & 0.051 \\
    Fig. 13(h) & 0.46 & 0.44 & 0.012\\
    Fig. 13(i) & 26.53 & 0.79 & 0.012 \\
    \hline
  \end{tabular}
  \caption{Fitted parameters \(a\), \(b\) and \(c\) for each part of Fig.~\ref{Fig12}}
  \label{table:fitted_parameters}
\end{table}

In the case of the Ising model, one can only be astonished about the convergence achieved in the renormalization process as the number of spins is increased. As we detail in  Appendix~\ref{app:T}, the reason behind this seems to be the effective closeness between the two operations that root at the renormalization procedure, i.e. the action of the renormalized  Hamiltonian, followed by the embedding operation, {\it vs} the embedding operation, followed by acting with the original Hamiltonian. The discrepancy between these two processes originate from the non isometric nature of the renormalization approximation. Otherwise, expectation values such as the magnitudes calculated in the paper would remain exactly the same after renormalization. In spite of this discrepancy, the difference between both operations, as evaluated from its squared norm, grows much slower than the squared norm of the Hamiltonian itself, which effectively brings them closer to an isometric behavior. We think that this explanation is behind the convergence which is observed in the renormalized magnitudes as the system size is increased.

\FloatBarrier
\section{Conclusions}
\label{sec:conclusions}

In conclusion, our analysis of the Ising model using the Block Renormalization Group Method (BRGM) has yielded valuable insights into the behavior of physical features such as magnetization, correlation function and entanglement entropy across different system sizes. The method implies a change in the number of spins by a factor $1/2$, which is a substantial reduction in practical terms, since the dimension of the associated Hilbert space scales as $2^N$. At variance with the original aim of the BGRM method, where the renormalization procedure is iterated infinitely to investigate the properties of critical points, the originality of our idea is to perform this process just once, thus allowing to compare the results obtained from the original Hamiltonian with those from the renormalized one. We observed that, as the value of N is increased, there is an exponential convergence between both Hamiltonians, provided that the coupling constants are redefined accordingly, within the spirit of the BRGM. In fact, for spin chains comprising 24 spins, all these features exhibited an exact correspondence with the results obtained with the original Hamiltonian. This remarkable similarity suggests that the BRGM successfully preserves, already at this size,  the essential physical characteristics of the Ising model while making use of just half the spins. We emphasize that the technique fails for small systems, so it is necessary that a sufficiently large problem is tackled when using this tool. 

We have also investigated the applicability of the BRGM to the Heisenberg model. For this model, the method entails a reduction in the number of spins by a factor $1/3$, which anticipates that achieving a good agreement between the calculated magnitudes from the original and renormalized Hamiltonians may require going beyond $N=24$. This is in fact observed in our calculations, where the difference between both cases certainly manifest a tendency to decrease. Unfortunately, a confirmation of convergence would require the access to computational capabilities that are beyond our present reach.

Although our calculations were performed on a classical computer, the techniques involved in the BGRM can be easily exported to a quantum computer. For the Ising model, they would allow obtaining exact results with one half the qubits needed to simulate the original Hamiltonian. We also expect that the same technique can be used for other Hamiltonians. Our results for the Heisenberg model also point towards this direction. Reducing the amount of required resources is of fundamental importance in the NISQ era, where quantum processors are severely constrained by their limited number of qubits.

The computation for the spin chain with 24 spins presented a significant computational challenge. To overcome this hurdle, we utilized the computational resources of the Lluis Vives machine at the Valencia University in Spain. The availability of such a powerful computing platform enabled us to accurately perform the necessary calculations and obtain reliable results for the 24-spin case.

Our findings will allow to  investigate the potential of BRGM to further models, and to extend it to higher dimensions.  This broadening of BRGM’s scope to encompass a wider range of Hamiltonians  holds the potential to propel our understanding of quantum systems and critical phenomena to new heights.

\section*{Acknowledgments}
The authors would like to acknowledge the enlightening discussions with Germán Sierra and express their gratitude to Ali Najafi (https://www.linkedin.com/in/ali-najafi-84345546) for his collaboration in numerical calculations. These calculations were performed using the Luis Vives machine at the University of Valencia, Spain. The authors gratefully acknowledge the computer resources at Artemisa, funded by the European Union ERDF and Comunitat Valenciana, as well as the technical support provided by the Instituto de Fisica Corpuscular, IFIC (CSIC-UV). This work has been funded by the Spanish MCIN/AEI/10.13039/501100011033 grant PID2020-113334GB-I00, Generalitat Valenciana grant CIPROM/2022/66, the Ministry of Economic Affairs and Digital Transformation of the Spanish Government through the QUANTUM ENIA project call - QUANTUM SPAIN project, and by the European Union through the Recovery, Transformation, and Resilience Plan: NextGenerationEU within the framework of the Digital Spain 2026 Agenda, and by the CSIC Interdisciplinary Thematic Platform (PTI+) on Quantum Technologies (PTI-QTEP+). This project has also received funding from the European Union’s Horizon 2020 research and innovation program under grant agreement 101086123-CaLIGOLA. 
M.Á.G.-M. acknowledges funding from the Spanish Ministry of Education and Professional Training (MEFP) through the Beatriz Galindo program 2018 (BEAGAL18/00203), QuantERA II Cofund
2021 PCI2022-133004, Projects of MCIN with funding from European Union NextGenerationEU (PRTR-C17.I1) and by Generalitat Valenciana, with Ref. 20220883 (PerovsQuTe) and COMCUANTICA/007 (QuanTwin), and Red Temática RED2022-134391-T. CGA acknowledges support from the Spanish Ministry of Science, Innovation and Universities through the Beatriz Galindo program 2020 (BG20-00023) and the European ERDF under grant PID2021-123627OB-C51 and from the QuantERA grant EQUIP with the grant numbers PCI2022-133004, funded by Agencia Estatal de Investigación, Ministerio de Ciencia e Innovación, Gobierno de España, MCIN/AEI/10.13039/501100011033, and by the European Union “NextGenerationEU/PRTR”

\appendix
\counterwithin{figure}{section}
\section{Discussion on convergence}
\label{app:T}
From the definition of \( P \),  and from Eq.~\eqref{Eq.4}, one finds
\begin{equation}
T H' = P H T \tag{A.1}.
\end{equation}
This suggests that the difference
\begin{equation}
\Delta H \equiv HT - TH' = HT - PHT \tag{A.2}
\end{equation}
can be used to quantify the error introduced by the renormalization approximation.
In fact, if \( \Delta H = 0 \), one has that
\begin{equation}
T H' = H T \tag{A.3}
\end{equation}
and this relation holds for any function $f$:
\begin{equation}
T f(H') = f(H) T \tag{A.4}.
\label{Eq:A4}
\end{equation}
Assume an initial state $|\psi_0\rangle$ in \(\mathcal{H}\), with corresponding $|\psi'_0\rangle = T^\dagger |\psi_0\rangle$.
After a time $t$ we have:
\begin{equation}
|\psi'(t)\rangle = e^{-iH't} |\psi'_0\rangle \tag{A.5}.
\end{equation}
Given an observable $A$ in \(\mathcal{H}\) and its corresponding renormalized version
\begin{equation}
A' = T^\dagger A T \tag{A.6},
\end{equation}
the expectation value on $|\psi'(t)\rangle$ is
\begin{equation}
\langle \psi'(t) | A' | \psi'(t) \rangle = \langle \psi'_0 | e^{iH't} 
 T^\dagger  A   T    e^{-iH't} | \psi'_0 \rangle  \tag{A.7}.
\end{equation} 
Using Eq.~\ref{Eq:A4}, one easily obtains that
\begin{equation}
\langle \psi'(t) | A' | \psi'(t) \rangle = \langle \psi(t) | A | \psi(t) \rangle  \tag{A.8},
\end{equation} 
with $| \psi(t)\rangle = e^{-iHt} | \psi_0 \rangle $.

This result was expected since, in this particular case, \( T \) represents an isometry between \(\mathcal{H}\) and \(\mathcal{H'}\).
The above reasoning suggests using the squared trace norm of \( \Delta H \)
\begin{equation}
\|\Delta H \|^2 = \mathrm{Tr} \{ (\Delta H)^\dagger \Delta H \} \tag{A.9}
\end{equation}
to quantify the discrepancy between both quantities \( TH' \) and \( HT \) as the dimension \( N \) of \(\mathcal{H}\) is increased. Notice, however, that $\|\Delta H \|$ can grow with \( N \), simply because the dimension of the Hilbert space increases. Hence, we use instead  the normalized quantity
\begin{equation}
\varepsilon(\Delta H) = \frac{\|\Delta H\|^2}{\| H \|^2} \tag{A.10}.
\end{equation}
After some algebra, one obtains
\begin{equation}
{\varepsilon}(\Delta H) = \frac{\mathrm{Tr} \{P H^2\} - \mathrm{Tr} \{(PH)^2\}}{\mathrm{Tr} \{H^2\}} \tag{A.11}.
\end{equation}
\FloatBarrier
\begin{figure}[ht]
    \centering
    \includegraphics[width=0.45\textwidth]{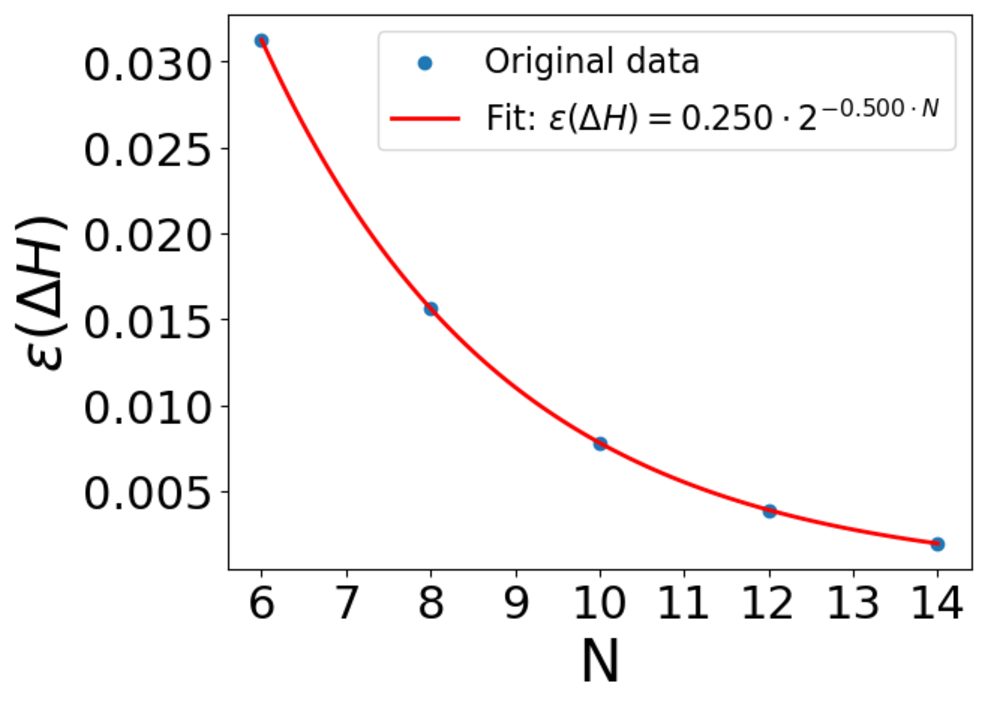}
    \caption{Plot of the normalized discrepancy \( \varepsilon(\Delta H) \) calculated for the Ising model, with $J=\Gamma=1$.}
    \label{Appendix}
\end{figure}
\FloatBarrier
\captionsetup{font=small}
The isometry case examined above appears when $ P = I_\mathcal{H} $, in which case both terms in the numerator of \( \varepsilon(\Delta H) \) are equal (they also coincide with the denominator), which further justifies the normalization by this operator. Figure \ref{Appendix} plots this magnitude as a function of the number $N$ of spins for the Ising model, with $J=\Gamma=1$. As can be seen, this magnitude goes exponentially to zero. In fact, it can be well fitted by the exponential $a 2^{-b N}$, using $a=1/4$, and $b=1/2$, similarly to the exponential behavior already found for calculated observables.

\bibliographystyle{unsrt}
\bibliography{Bibliography}

\end{document}